# Features of the flexomagnetoelectric effect in an external magnetic field.

**Vakhitov R.M., Solonetskiy R.V., Nizyamova A.R.**

[1]Bashkir State University, 450076, Ufa, Russia

vakhitovrm@yahoo.com

The paper investigates the influence of the magnetic field on the behavior of 180-degree domain boundaries in a uniaxial ferromagnetic film with inhomogeneous magnetoelectric interaction. It is shown that, depending on the magnitude and direction of the field, it is possible to strengthen or weaken the flexomagnetoelectric effect in the sample under study. In addition, it was found that in the reverse field, the effect of switching the nature of the interaction of the electric field source with the domain wall from attraction to repulsion is possible.

## INTRODUCTION

Currently, studies of magnetoelectric effects observed in a certain class of magnets called multiferroics are of increased interest [1,2]. They are characterized by two or more order parameters and have a number of unusual properties that can be used in spintronics and magnetic memory devices of a new generation. Multiferroics, as is known, also include ferrite-garnet films, in which a giant magnetoelectric effect (linear) was detected at room temperature [3]. After some time, a new effect of this type was discovered in them, consisting in the phenomenon of displacement of domain boundaries (DG) under the action of an inhomogeneous electric field [4]. Analyzing the experimental data, the authors suggested that they can be explained by the manifestation of the flexomagnetoelectric effect (FME) [1], i.e., the presence of inhomogeneous magnetoelectric interaction (NMEV) in the studied materials, first considered in [5]. The results obtained in [4] initiated new research in this direction [6-11], which made it possible to study more thoroughly the effect of the electric field on the structure and properties of magnetic inhomogeneities of various topologies in magnetic films with NMEV. At the same time, another interpretation of the experimental data [3] was proposed in [12,13], which is not related to "charged" DGS. It is based on the effect of a possible change in the value of the anisotropy constant of the material due to the displacement of the same type of ions relative to the equilibrium position under the action of an inhomogeneous electric field. It should be noted that in [14] the flexomagnetoelectric nature of induced electric polarization in ferrite-garnet films was confirmed on the basis of fluorescence spectroscopy of single molecules. Nevertheless, a comparative analysis of the above mechanisms showed [15] that both of them at a qualitative level fully explain the picture of the behavior of DG in an inhomogeneous electric field. It follows that each of the mechanisms contributes to the phenomenon under study. However, which of them is the dominant one will need to be found out in the course of further research. In addition, it is of practical interest to study various factors (external or internal) that significantly affect the degree of manifestation of this effect. In particular, it was shown in [15-19] that some properties of the DG (the magnitude of its displacement, its velocity, etc.), as well as its transformation in an inhomogeneous electric field, are significantly influenced by an external magnetic field, and, in particular, its planar component [15,19]. To this end, this paper provides a theoretical analysis of the influence of an external magnetic field on the nature of the manifestation of FME in the studied magnets.

## PROBLEM STATEMENT

A uniaxial ferromagnet in the form of a film with a thickness of D. is considered. It is assumed that the axis of light magnetization of perpendicular anisotropy is directed along the normal to the film and parallel to the Oz axis (Fig.1), and the Oy axis coincides with the direction along which the sample it is inhomogeneous, i.e. magnetic moments rotate along it. The magnetization vector M=$M_s$ m ($M_s$ is the saturation magnetization) is expressed in terms of the unit vector m, defined through the variables $\theta$ and $\varphi$: m=(sin$\theta$cos$\varphi$,sin$\varphi$,cos$\theta$cos$\varphi$).

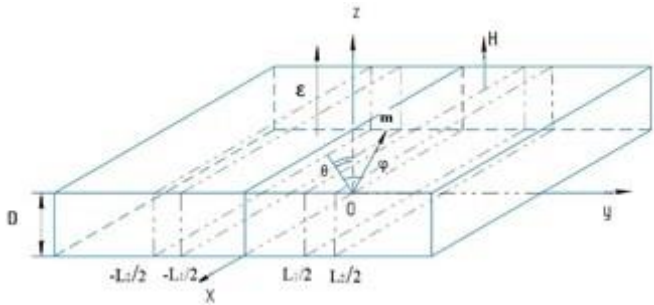

Fig.1 Diagram illustrating the geometry of the problem

The energy of the magnet reduced to the cross-sectional area of the film by the xOz plane is taken as:

$$E= \int_{-\infty}^{\infty} \left\{A\left[\left(\frac{d\varphi}{dy}\right)^2 + \cos^2\varphi \left(\frac{d\theta}{dy}\right)^2\right] + K_u(\sin^2\theta\cos^2\varphi + \sin^2\varphi) + \varepsilon_{int} + \varepsilon_H + 2\pi M_s^2 \sin^2\varphi\right\} dy , \quad (1)$$

where A is the exchange parameter, $K_u$ is the uniaxial anisotropy constant, $e_{int}$, $e_H$ are the energy densities of the NMEV and the Zeeman interaction, respectively, and the last term represents the energy density of demagnetizing fields from bulk charges [20,21]. It is assumed that the film is thick ($\Delta_0 \ll D < \Lambda_0$, $\Delta_0 = \sqrt{A/K_u}$ is the characteristic size of the DW , $\Lambda_0 = \sqrt{A/2\pi M_s}$ is the size of the Bloch line [22]) and the contribution of demagnetizing and scattering fields is neglected. Accordingly, the formula for $\varepsilon_H$ has the form:

$$\varepsilon_H = -M_s(\mathbf{mH}) \qquad (2)$$

and the expression for $\varepsilon_{int}$ is taken in the form [23]

$$\varepsilon_{int} = M_s \mathcal{E}(b_1 \mathbf{m}\,\mathrm{div}\,\mathbf{m} + b_2 \mathbf{m}\,\mathrm{rot}\,\mathbf{m}) \qquad (3)$$

where b1, b2 are magnetoelectric constants, $\mathcal{E}$ and $\mathbf{H}$ are the strengths of the electric and magnetic fields, respectively. In this case, these fields are considered inhomogeneous and act in bounded regions of space,

$$\mathcal{E} = \mathcal{E}_0/\mathrm{ch}^{-1}(y/L_1), \quad H = H_0/\mathrm{ch}^{-1}(y/L_2), \qquad (4)$$

where $\mathcal{E}_0 = \mathcal{E}(0)$, $H_0 = H(0)$ are the values of the corresponding fields in the center of their action band, $L_1, L_2$ are the characteristic sizes of the corresponding bands along the Oy axis. It is assumed that the field $\mathcal{E}$ is directed along the axis Oz (E||Oz), and the field $\mathbf{H}$ is arbitrary.

Then, the expression for $\varepsilon_{int}$, written through angular variables, will take the form:

$$\varepsilon_{int} = \mathcal{E}M_s^2 \left[(b_1\cos^2\varphi + b_2\sin^2\varphi)\cos\theta\frac{d\varphi}{dy} + b_2\sin\theta\sin\varphi\cos\varphi\frac{d\theta}{dy}\right] \qquad (6)$$

The structure and properties of magnetic inhomogeneities are determined from the Euler-Lagrange equations, which have the form:

$$\frac{d}{d\xi}\left(\frac{\cos^2\varphi\,d\theta}{d\xi}\right) - \sin\theta\cos\theta\cos^2\varphi + \frac{(\lambda_1 + \lambda_2)f(\xi)\sin\theta\cos^2\varphi\,d\varphi\,d\varphi}{d\xi}$$

$$+ \lambda_2 \sin\theta\sin\varphi\cos\varphi\frac{df(\xi)}{d\xi} - \frac{\frac{\partial\varepsilon_H}{\partial\theta}}{M_s H_u}$$

$$= 0 \qquad (5)$$

$$\frac{d^2\varphi}{d\xi^2} - \sin\varphi\cos\varphi\left[\cos^2\theta - \left(\frac{d\theta}{d\xi}\right)^2\right] - (\lambda_1 + \lambda_2)f(\xi)\sin\theta\cos^2\varphi\frac{d\theta}{d\xi} + (\lambda_1\cos^2\varphi + \lambda_2\sin^2\varphi)\cos\theta\frac{df(\xi)}{d\xi} - \frac{1}{M_s H_u}\frac{\partial\varepsilon_H}{\partial\varphi} - Q^{-1}\sin\varphi\cos\varphi = 0,$$

where $\lambda_i = \mathcal{E}_0/\mathcal{E}_i = \mathcal{E}_0 M_s^2 b_i/2K_u\Delta_0$, $\mathcal{E}_i = 2K_u\Delta_0/M_s^2 b_i$, $i = 1, 2$; $\xi = y/\Delta_0$, $l_i = L_i/\Delta_0$, $f(\xi) = \mathrm{ch}^{-1}(\xi/l_i)$, $Q = K_u/2\pi M_s^2$. Here $\lambda_i, \mathcal{E}_i$ – accordingly, the following and characteristic electric fields, ξ is the reduced coordinate, Q is the material quality factor, $H_u = 2K_u/M_s$ is the uniaxial anisotropy field. In the future, another dimensionless parameter h=$H_0$/$H_u$ (the reduced magnetic field) will be involved.

Numerical analysis of these equations taking into account NMEV showed [21] that in uniaxial ferromagnets at h=0, depending on the selected boundary conditions imposed on θ and φ at |ξ|→∞, the existence of three types of micromagnetic structures is possible. These are 180° DW with a non-circular trajectory of the magnetization vector [24], 0° DW with a quasi-Bloch structure [24,25], 0° DW of non-Aelian type [25]. In this paper, the main attention will be paid to the behavior of 180° DW in the ferromagnet under study in an external magnetic field, which is associated with similar experimental studies of FME [15,19], in which only this type of boundaries was observed.

### 3. Transformation of 180° DG in an electric field, h=0

Obviously, a 180° DW of the Bloch type will transform in the external magnetic field **H**, but the nature of these changes will depend on both the magnitude and orientation of the field **H**, relative to the plane of the DW. At the same time, the case when **H**||Oz, for 180° DW does not make sense to consider, because such a field will only lead to the displacement of DW as a whole.

At the beginning, consider the case h=0. Numerical investigation of equations (5) (here we consider the case λ1=λ2= λ) shows [21] 180° DW Bloch-type under a non-uniform electric field undergoes a series of transformations of its texture with increasing values of λ: 180° Bloch-type DW →180°DW with a quasi-Bloch structure → 180° DW with a quasi-Neelian structure →180° DW of non-Aelian type. Magnetic inhomogeneities located in this chain of transformations in intermediate positions belong to the DW with a non-circular trajectory of the magnetization vector [10,21,24]. This means that the magnetic moments in both types of DW have both Bloch (mx ≠ 0) and non-Nobel (my ≠ 0) components. However, their difference lies in the fact that 180° DW with a quasi-Bloch structure does not have sections with a purely non-Nobel law of rotation of magnetic moments (mx = 0), and in the second type there are such sections.

It should be noted that a cascade of transformations of the structure 180° DW occurring with increasing electric field, accompanied by the first induction in the area 180° DW related charges, and a subsequent increase in the electric polarization (as it differential value of p=ν$p_0$ and the integral P=N$p_0$, where ν and N are given, respectively, differential and integral polarization $p_0 = M_s^2 b_i \Delta_0$-characteristic value of polarization [21]). When the field reaches the value λ=λc, at which 180° DW becomes completely non-Abelian, on the graph of the dependence N= N(λ) (Fig. 2, black curve) there is a break: a sharp rise is replaced by a section of a slow (adiabatic) increase in the magnitude of N.

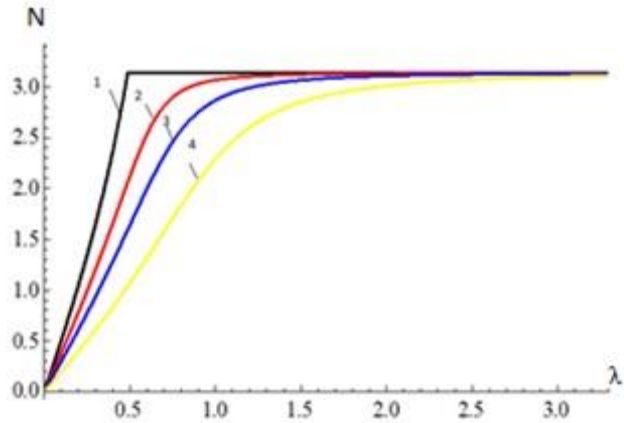

Fig.2 Dependences of the integral polarization value N 1800 DG on the parameter λ in the magnetic field **H** || Ox. Line 1 (black) corresponds to x = 0, red - h = 0.1, blue - x = 0.2, yellow H = 0.4. Here and in the future, the values of the material parameters are taken as follows: M = 3, $L_1 = 5$, $P_2 = 1000$.

### 4. Transformation of 180° DG in a magnetic field, λ=0

Consider the effect of an external magnetic field on the structure and properties of 180° DW. Assume that **H**||Ox and coincides with the direction of the magnetic moments in the plane DG at y = 0. In this case, the magnetic moments form an angle with the field ψ, lying in the range $0 \leq \psi \leq \theta_0$, where $\theta_0 =$ arcsin(h). The analysis of equations (6) for this case shows that in the absence of an electric field (λ=0), the magnetization in the domains **M₀** is with the axis Oz angle $\theta_0$. Accordingly, 180° DW of the Bloch type under the action of a magnetic field h becomes narrower (180-2$\theta_0$) – degree with the law of rotation of the vector **m** in the wall, defined by the expressions (at $l_2 \to \infty$).

$$\theta = 2\text{arctg}\{[1 - \sqrt{1-h^2}\,\text{th}\,(\sqrt{1-h^2}\,\xi/2)]/h\}, \varphi = 0 \qquad (7)$$

From this it can be seen that as the field h increases, the maximum angle of rotation of the magnetization $\theta_m$ in such a DW, equal to $\theta_m = (180-2\theta_0)°$, will continuously decrease, and its width Δ will increase (Fig. 4).

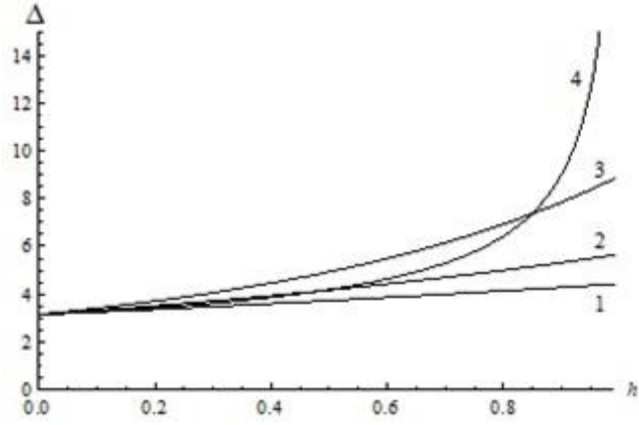

Fig.3 Dependences of the width of $180^0$ DW $\Delta$ on the magnetic field ($\mathbf{H}\|Ox$).Here $\lambda = 0$, line 1 corresponds to $l_2 = 3$, 2 - $l_2 = 5$, 3 - $l_3 = 10$, 4 - $l_2 = 1000$.

When the critical value h=1 ($H=H_u$) is reached by the field h, the limiting orientations of magnetization in the domains $\mathbf{m_1}$ and $\mathbf{m_2}$ ($\mathbf{m_1}=\mathbf{m}(-\infty)$,$\mathbf{m_2}=\mathbf{m}(\infty)$) become parallel ($\mathbf{m_1}\|\mathbf{m_2}$), and the width of such a DW increases indefinitely. Accordingly, $\theta_m \to 0$ and the wall disappears. However, if the magnetic field is inhomogeneous and acts in a limited area representing a strip of width $l_2$ (along the Oy axis), then in this case, according to calculations, with increasing h, the width of the DW $\Delta$ will also increase, but with a smaller angle of inclination of the corresponding curve (Fig.3). At the same time, the U-turn angle $\theta_m$ will decrease, but it will reach the limit value $\theta_m=0$ in much larger fields (h>1).

If the magnetic field is directed opposite to the Ox axis, then 180° DW will transform according to a different scenario. In this case, the magnetic moments in the domains they will also begin to deviate from the Oz axis towards the direction of the field $\mathbf{H}$, but the reversal of the vector $\mathbf{m}$ will already be $\theta_m \geq \pi$. The structure 180° DW will be described in a different distribution of the magnetization, which has the form (when $l_2 \to \infty$)

$$\theta = -2\mathrm{arctg}\{[1 + \sqrt{1-h^2}\mathrm{cth}(\sqrt{1-h^2}\xi/2)]/h\}, \varphi = 0 \qquad (8)$$

Accordingly, the hodograph of the magnetization vector $\mathbf{m}$ will describe a longer trajectory on the surface of a sphere of unit radius ($\theta_m=\pi+2\theta_0$) than in the first case of orientation $\mathbf{H}$. Thus, this wall represents $(180+2\theta_0)°$ DW. As h increases, the angle θm will also increase in the limit at h=1 angle $\theta_m=2\pi$, i.e. $(180+2\theta_0)°$ DW will become 360°DW. Accordingly, the magnetic moments located in the center of the wall (near y = 0) will be directed opposite to the field $\mathbf{H}$. As is known, such a wall becomes unstable with respect to fluctuations of the magnetization vector of the non-Abelian type and at a certain value of the field h [26] collapses and disappears

If the magnetic field H is directed along the Oy axis, then there is a qualitative change in the structure of 180 ° DG in the magnetic field. In this case, the wall, while remaining 180-degree, is transformed from a Bloch type to a quasi-Bloch wall, since the magnetization M exits the plane DW ($\varphi \neq 0$). In addition, the magnetization in the $\mathbf{M_0}$ domains deviates from the xOz plane (coinciding with the DW plane) by the angle $\varphi_0 = \varphi(\infty) \neq 0$ (Fig.4 green dashed line (1')).

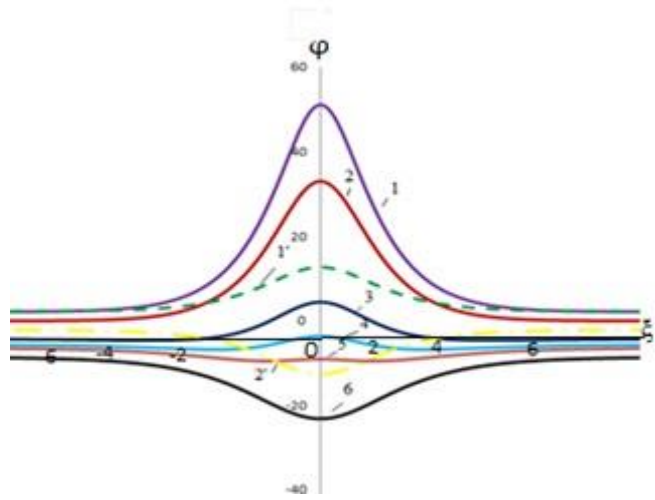

Fig.4 Dependences of the angle $\varphi$ on the given coordinate $\xi$ in the magnetic field H||Oy for different values of h. Here $\lambda=0$, line 1' (green dashed) corresponds to h=0.1, 2' (yellow dashed)-h=-0.1, $\lambda$ =0.3:line 1 (purple) corresponds to the value h = 0.1, line 2 (red) - h = 0, line 3 (green) - h= -0.2, line 4 (blue) - h = -0.26, line 5 (yellow) - h= - 0.3, line 6 (black) - h= -0.4

With increasing h, the maximum exit angle $\varphi_m$ increases and at some h=$h_1$ reaches values of $\varphi_m=\pi/2$. With a further increase in the field up to h=$h_2$ (at Q = 3, $l_2$ = 1000, $h_2$=0.4), the non-Abelian contribution to the structure of DW increases (my increases), and the Bloch contribution decreases ($m_x \to 0$). Finally, at h=$h_2$, the wall becomes completely non-Gel. The subsequent increase in h causes the wall to become unstable and collapse. As the size of the inhomogeneity band of the magnetic field $l_2$ decreases, this critical field increases. In the reverse field, the transformation process of 180° DW is completely repeated, but the angle $\varphi_m$ in this case will take the values of the opposite sign.

## 5. Conversion of 180° DW in a magnetic field, ($\lambda \neq 0$)

Let us now study the influence of an external magnetic field on the flexomagnetoelectric effect. We assume that H||Ox, and the chirality of DW is such that the magnetic moments (at y=0) coincide with H. Then, when the field is "turned on", a similar transformation will take place, discussed in the previous section: 180° DW with a quasi-Bloch structure is transformed into (180-2$\theta_0$)° DW also with the exit of m from the plane of rotation of magnetic moments (Fig.5a). However, at the same time, with an increase in the magnitude of h, which tends to rotate the magnetic

moments along the field (i.e. to return them again to the DW plane with the constant value of the parameter λ), the maximum exit angle $\varphi_m$ decreases (Fig.5b).

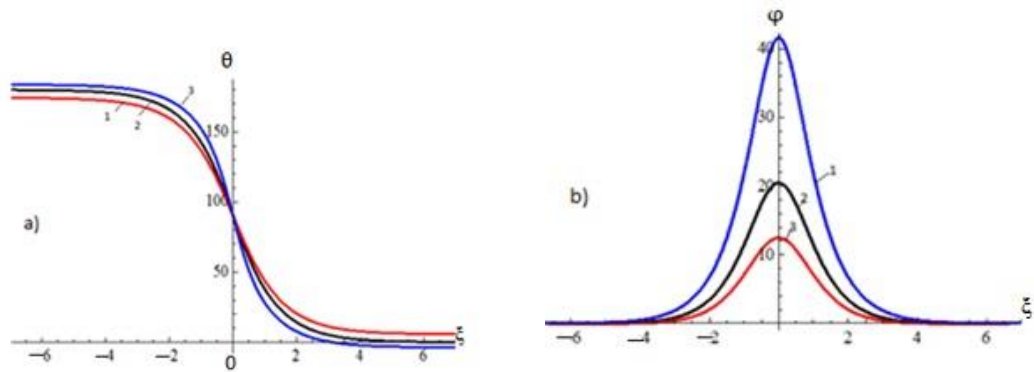

Fig. 5. 180° DG profiles determined by the dependencies of the angular variables $\theta$ (a) and $\varphi$ (b) on the reduced coordinate $\xi$ in the magnetic field H||Ox: line 1 (black)- h = 0, line 2 (red) -h = 0.2, line (3 blue) h = -0.13. Here $\lambda$ = 0.2.

In addition, the maximum value of the differential polarization $p_m$ also decreases (Fig. 6).

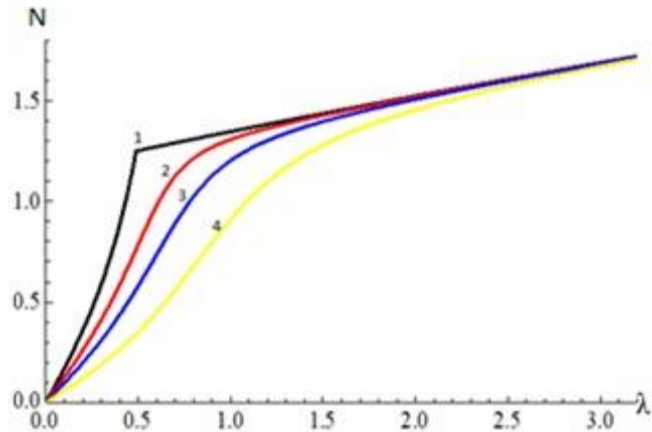

Fig.6 Dependences of the value of $v_m$ for 180° DW on the parameter $\lambda$ in the magnetic field H||Ox. The black line corresponds to h = 0, red - h = 0.1, blue - h = 0.2, yellow h = 0.4.

The latter leads to a decrease in the value of the integral polarization N. However, with the increase of the electric field $E_0$ (increasing $\lambda$) is the relative value of decreasing $\Delta N/N (\Delta N=N(h_2)-N(h_1), N$ ($h_i$)-values of the integral polarization considered for different values of the $h_i$ (i=1.2), but with the same value of $\lambda$) will decrease, while in the limit reaches zero. In this case, all the dependence curves N=N($\lambda$) converge in the limit ($\lambda \to \infty$) to the same asymptote (Fig.2), which corresponds to the dependence curve 180° DW of non-Abelian type (h=0). The same behavior is demonstrated by the dependence curves $\varphi_m = \varphi_m (\lambda)$ and

$v_m = v_m(\lambda)$. It follows that the effect of a magnetic field with H∥Oh weakens the FME. In addition, the presence of a magnetic field leads to a smoothing of the transition from 180° DW of the quasi–Parallel type to 180° DW of the Neel type (on the graphs of the dependence $N=N(\lambda)$), (Fig.2) there is no "break" of the curves from λ), and also to a decrease in the critical field $\lambda c$ of such a transition.

If the magnetic field is directed opposite to the Oh axis, then the magnetic moments in the domains will deviate from the Oz axis in the opposite direction and 180° DW will also be converted into $(180-2\theta_0)°$ DW. At the same time, the magnetization exit angle from the wall plane increases significantly (Fig.6), respectively, the differential polarization v increases, which leads to an increase in the integral polarization N (Fig.3). Thus, in the reverse field, the FME in the sample under study increases significantly.

Let us now consider the situation when the magnetic field H∥Oy acts on the initial magnet. In this In the case already at $\lambda=0$ 180 ° DW of the Flea type, it is transformed under the action of a magnetic field into a quasi-Bloch wall. In this case, the magnetization in the domains $\mathbf{M_0}$ deviate from the xOz plane by an angle $\varphi\_0$ (Fig.4). At $\lambda \neq 0$, the process of changing the topology of the wall increases; with increasing magnitude h, the angle $\varphi\_0$ and the maximum angle of deviation from the homogeneous state $(\varphi_m - \varphi_0)$ increases, and the maximum value of the differential polarization $v_m$. Accordingly, the value of the integral polarization N increases (Fig. 7).

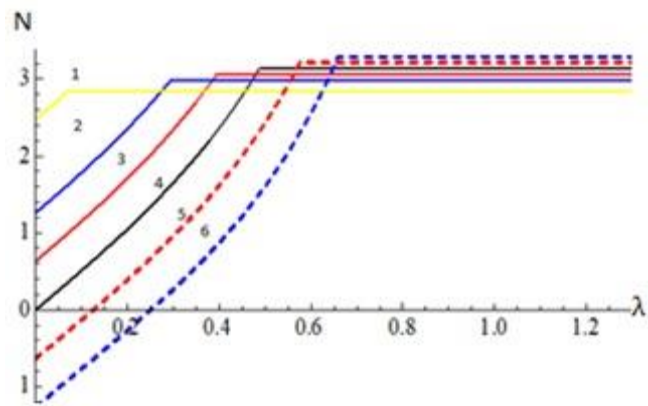

Fig.7 Dependences of the integral polarization of 180° DW on the parameter λ in the magnetic field H ∥ Oy. The black line corresponds to p = 0, red - p = 0.1, blue - p = 0.2, yellow - p = 0.4, red dashed - p = -0.1, blue dashed - p = -0.2.

At the same time, an interesting pattern is observed: the higher the value of h, the lower the electric fields, the transition of the quasi-Bloch 180° DW into the non-Nobel wall is achieved, at the same time the maximum value of the integral polarization decreases (Fig.8).

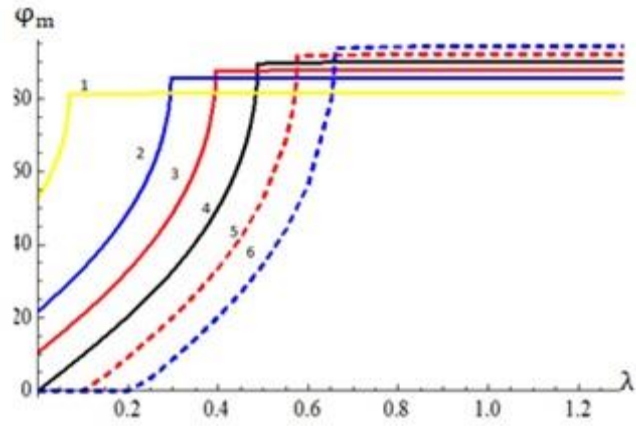

Fig.8 Dependences of the maximum angle of deviation from the homogeneous state $\varphi_m$ 180° DW on the parameter $\lambda$ in the magnetic field H || Oy. The black line corresponds to p = 0, red - p = 0.1, blue - p = 0.2, yellow - p = 0.4, red dashed - p = -0.1, blue dashed - p = -0.2.

When the electric field reaches its critical value $\lambda=\lambda c$, the structure of 180° DW becomes non-Neel. At the same time, the graph shows the dependence of the magnitude of the integral polarization N from $\lambda$ also has a fracture, similar to what it was at h=0. It follows that under the action of the magnetic field H along the Oy axis, the FME increases, but this happens in small fields h, and in large fields the effect weakens.

In the case when the direction H is opposite to the axis Oy, the magnetic moments, turning towards the field, as a result form an angle $\varphi_0 = \varphi(\infty)$, which becomes negative and lowers the maximum angle of magnetization exit from the plane DW $\varphi_m$ (Fig.4). As a result, the values of $v_m$ and N decrease. With a further increase in h, the value of N decreases and at a certain value h = $h_0$, it becomes zero, and at h> $h_0$ - negative (Fig.8). This means that 180° DW will have to repel from the source of an inhomogeneous electric field. Thus, by switching the direction of the magnetic field, it is possible to change the sign of the polarization magnitude and thereby change the nature of the interaction of 180° DW with an external electric field. The obtained result is in good agreement with experimental data [4]. It allows using electric and magnetic fields to regulate the movement of the DW, which is of practical interest.

## Discussion of the results

Thus, it follows from the above results that the presence of an external magnetic field has a significant effect on the flexomagnetoelectric effect observed in films of ferrite garnets with NMEV. The degree of its impact depends on both the magnitude and the orientation of the magnetic field relative to the plane of 180° DW. In particular, in this paper, the structure of 180° DW was studied with its two mutually perpendicular directions: **H**||Ox, **H**||Oy. According to calculations, a significant (multiple) amplification of the effect will take place when electric and magnetic

fields in the following geometry act on 180 ° DW: E ||Oz, **H** || Oy, and the greatest amplification effect can be achieved already in small magnetic fields. This is consistent with experimental data [15, 19], from which it follows that the greatest displacement of the DW in an inhomogeneous electric field occurs when a magnetic field perpendicular to the wall plane acts. In this case, the effect of increasing the value of the integral polarization N is achieved by increasing the magnitude of the exit angle of the magnetization vector from the DW plane. Accordingly, the magnitude of volumetric magnetic charges increases, determined by the expression $\rho_v = -M_s \text{div}\mathbf{m}$ [18, 26], which ultimately leads to an increase in the parameters $v_m$ and N.

It also follows from the results obtained that by changing the orientation of the magnetic field to the opposite, it is possible to change the nature of the flexomagnetoelectric effect: either to strengthen it (in the case of **H**||Ox), or to weaken it. However, by switching the direction **H**, it is also possible to achieve a change in the nature of the interaction of the DW with the electric field from the attraction of the DW to its repulsion and vice versa. This property may be important in applied development. On the other hand, this property indicates that the flexomagnetoelectric mechanism is dominant even when exposed to an inhomogeneous electric field on the DW. The fact is that a perpendicular magnetic field can change the width of the DW, its topology, but not move it.

The work was carried out with the financial support of the State Task for the performance of scientific research by laboratories (Order MN- 8/1356 of 09/20/2021)